\documentclass[draft,grl]{AGUTeX}

\usepackage{graphicx,color,ulem}
\usepackage{sidecap}

\authorrunninghead{LINCOT ET AL.}

\titlerunninghead{Inner core made of cubic iron?}

\begin{document}

%
%

%
%

\authoraddr{Ainhoa Lincot, ISTerre, Universit\'e de Grenoble Alpes, BP 53, 38041 Grenoble CEDEX 9, France. }
\authoraddr{S\'ebastien Merkel, UMET, Universit\'e Lille 1, 59655 Villeneuve d'Ascq, France.}
\authoraddr{Philippe Cardin, ISTerre, Universit\'e de Grenoble Alpes, BP 53, 38041 Grenoble CEDEX 9, France. (philippe.cardin@ujf-grenoble.fr)}

\title{{Is inner core seismic anisotropy a marker of plastic flow of cubic iron?}}

\authors{A. Lincot,\altaffilmark{1,2}, S. Merkel,\altaffilmark{2}, P. Cardin,\altaffilmark{1}}

\altaffiltext{1}{Institut des Sciences de la Terre (ISTerre), Universit\'e de Grenoble Alpes, CNRS, Grenoble, France.}

\altaffiltext{2}{Unit\'e Mat\'eriaux et Transformations (UMET), ENSCL, CNRS, Universit\'e Lille 1, Villeneuve d'Ascq, France.}

%
%


\begin{abstract}
This paper investigates whether observations of seismic anisotropy are compatible with a cubic structure of the inner core Fe alloy.
We assume that anisotropy is the result of plastic deformation within a large scale flow induced by preferred growth at the inner core equator.
Based on elastic moduli from the literature, bcc- or fcc-Fe produce seismic anisotropy well below seismic observations ($<0.4\%$).
A Monte-Carlo approach allows us to generalize this result to any form of elastic anisotropy in a cubic system.
Within our model, inner core global anisotropy is not compatible with a cubic structure of Fe alloy.
Hence, if the inner core material is indeed cubic, large scale coherent anisotropic structures, incompatible with plastic deformation induced by large scale flow, must be present.
\end{abstract}

%
%

%

\begin{article}

%
%

\section{Introduction}


Seismic observations provide strong evidence that the Earth's inner core is anisotropic, with larger velocity in the polar than in the equatorial direction \citep[see][for recent review]{deuss2014areps}. This anisotropy is observed both using short wavelength body waves differential travel times \citep[e.g.][and references therein]{Irving11a}, and long period normal modes \citep[e.g.][and references therein]{Irving11b}. Over the years, this observation was refined with further evidences for both hemispherical \citep[e.g.][and references therein]{Tanaka12} and radial variations of the anisotropy amplitude and geometry, with an almost isotropic layer at the top of the inner core surrounding a more anisotropic region \citep{Souriau03} and possibly an innermost inner core with different properties \citep[e.g.][and references therein]{Lythgoe14}.


This anisotropic structure is likely due to an alignment of anisotropic Fe-alloys acquired either during solidification \citep[e.g.][]{Bergman97} or by deformation afterwards [\citet{Yoshida96}, and \citet{deguen2012epsl} for a recent review], although other hypothesis have been proposed, such as the presence of melt inclusions in the solid inner core \citep{Singh00}. The stable form of inner core  Fe-alloy remains a matter of debate: hexagonal {close-packed} (hcp), body-centered cubic(bcc) and face-centered cubic (fcc) have been proposed in the literature. Experiments imply that the stable form for pure Fe at inner core conditions is hcp  \citep[e.g.][]{Tateno10} but first-principles calculations also proposed various structures [bcc, \citet{belonoshko2008science}, fcc, \citet{mikhaylushkin2007prl}, or hcp, \citet{modak2007jpcm}]. However, differences in the free energy of the various Fe polymorphs
are very close to each other \citep{bouchet2013prb}. Moreover, the effect of impurity elements such as Ni or Si is still not well constrained \citep[see][for a recent review]{morard2014crg}. As such, a cubic structure for Fe at inner core conditions
remains a working hypothesis.

\begin{figure}[b!]
\centering
\includegraphics[width=7cm]{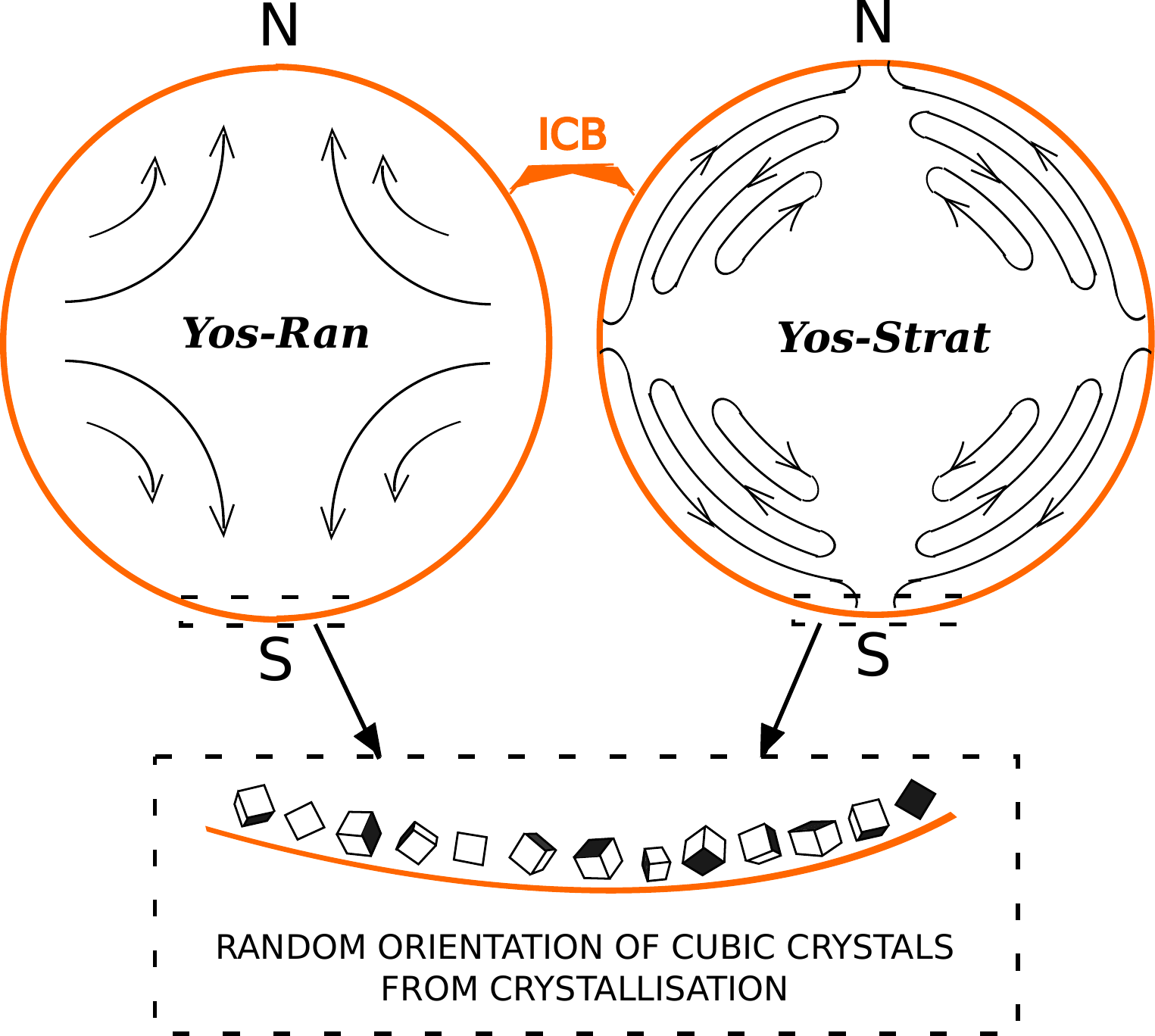}
\caption{Inner core geodynamical models used in this study. In \textit{Yos-Ran}, a quadripolar flow  results from the preferred equatorial growth of the inner core \citep{Yoshida96}. \textit{Yos-Strat} adds the contribution of a stable density stratification, resulting in layered flow pattern \citep{Deguen09}.}
\label{figmodele}
\end{figure}

Recently, we developed a framework for scaling microscopic properties such as single-crystal elasticity to the macroscopic observations of anisotropy \citep{Lincot14}. The model allows us to build a synthetic inner core with a given choice of crystal structure (previously hcp), elastic moduli, deformation mechanisms, and geodynamical model. In a second stage, we simulate the observations of body wave differential travel times that can then be compared with seismic observations.

Here, we investigate the effect of a cubic structure for Fe in such model and demonstrate that, within our framework, a cubic structure for Fe at inner conditions fails to reproduce observations of seismic anisotropy in the inner core.

\section{Methods} \label{part1}


As in \citet{Lincot14}, our geodynamical model is based on that of \citet{Yoshida96} (Fig.~\ref{figmodele}). The model assumes that geostrophic convection in the outer core induces faster crystallization of the inner core in its equatorial belt. The resulting topography is continuously relaxed by a quadrupolar flow, generating a plastic deformation ({model} \textit{Yos-Ran}, Fig.~\ref{figmodele}). We also use the extension of \citep{Deguen09} that accounts for a stable density stratification during inner core formation. Such model localizes the motions in the outer portion of the inner core, allowing much larger deformations ({model} \textit{Yos-Strat}, Fig.~\ref{figmodele}).

Our choice of geodynamical model is driven by the objective of producing lattice preferred orientations (LPO) aligned with the Earth's rotation axis, leading to a large North-South seismic anisotropy. To this end, quadrupolar models, such as \textit{Yos-Ran} and \textit{Yos-Strat}, are most efficient. Other models with smaller scale inner core deformation can result in strong LPO at the local scale and, hence, a large local anisotropy. However, such complex velocity structures within the inner core are averaged over the path of a seismic ray and will fail to produce global "average" anisotropy at the inner core scale \citep{Lincot14}.

Our study focuses on anisotropy acquired by deformation and does not account for anisotropy acquired during crystallization. {However,} crystallization
textures in cubic materials are complex and, hence, will probably not lead to any significant global scale anisotropy. Similarly, non axisymmetric models, such as those involving thermal translation of the inner core, are out of the scope of this study, which focuses on models preserving a symmetry
around the axis of rotation of the Earth.

\begin{figure*}
\centering
\includegraphics[width=12cm]{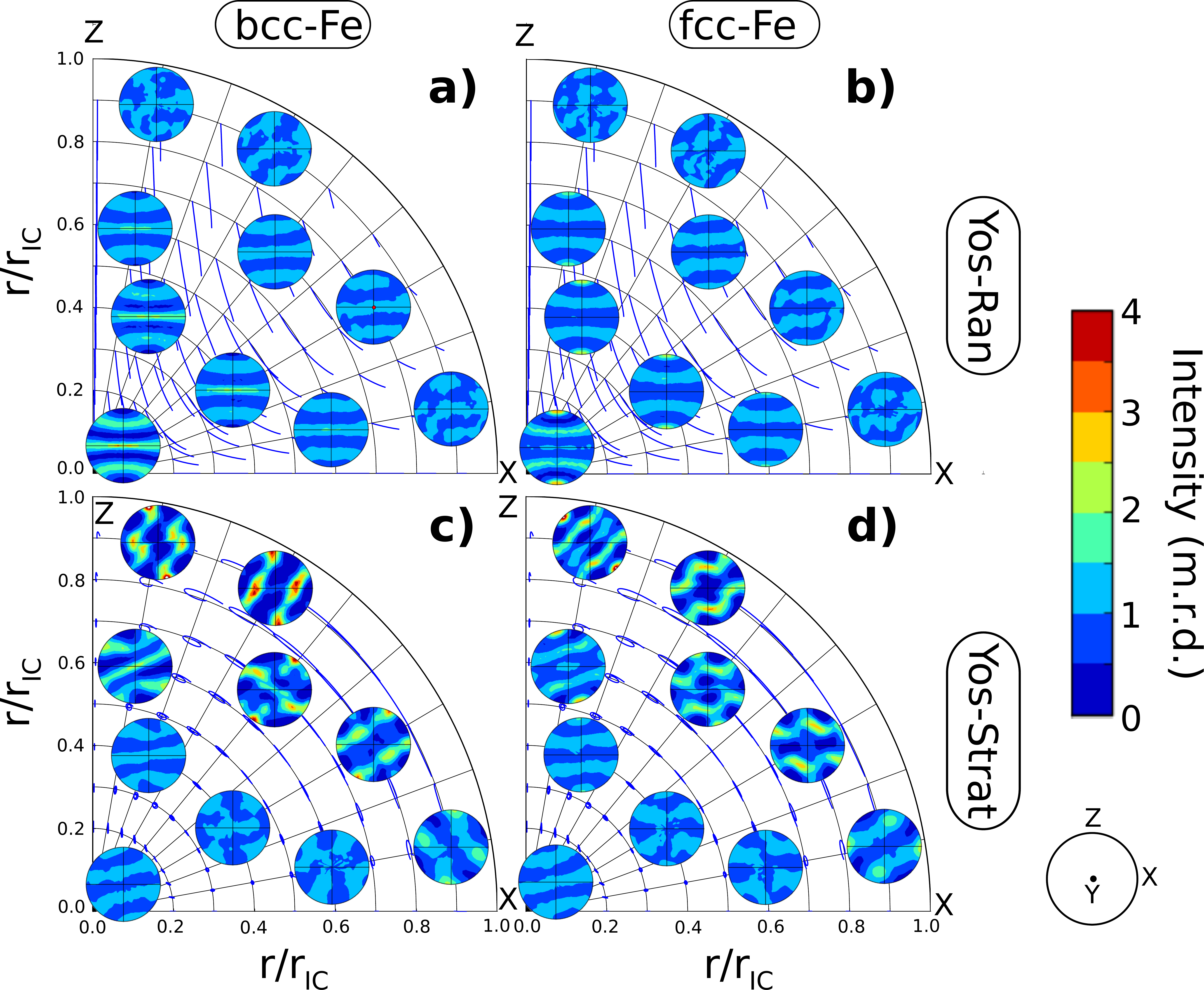}
\caption{{\it 100} pole figures representing present-day textures in Fe aggregates at various locations in the inner core for models \textit{Yos-Ran} (\textit{a,b}) and \textit{Yos-Strat} (\textit{c,d}) and either bcc-Fe  (\textit{a,c}) or fcc-Fe  (\textit{b,d}). The vertical $z$ axis is the geographical South-North axis while $x$ lies in the equatorial plane. Both geodynamical models are symmetric about $x$ and axisymmetric around $z$. Blue lines are the trajectories of the polycrystalline aggregates after crystallization at the ICB. Contours for pole figures in multiples of a random distribution (m.r.d.).}
\label{figTextures}
\end{figure*}

As in \citet{Deguen11a} and \citet{Lincot14},  we compute the position and deformation for 100 markers introduced at the top of the inner core boundary (ICB) during inner core growth (Fig.~\ref{figTextures}). Texture along these markers are calculated for a 3000 grains aggregate of cubic Fe using the Los Alamos viscoplastic self-consistent (VPSC) code  of \citet{Lebensohn93}, assuming a random crystallization texture and dominant slip along $\{111\}\langle \overline{1}10\rangle$ and $\{110\}\langle 1\overline{1}1\rangle$ for the fcc and bcc structure, respectively \citep{kocks1998}. Computed textures for a present day in core are presented in Fig.~\ref{figTextures}.

\begin{figure}
\centering
\includegraphics[width=7cm]{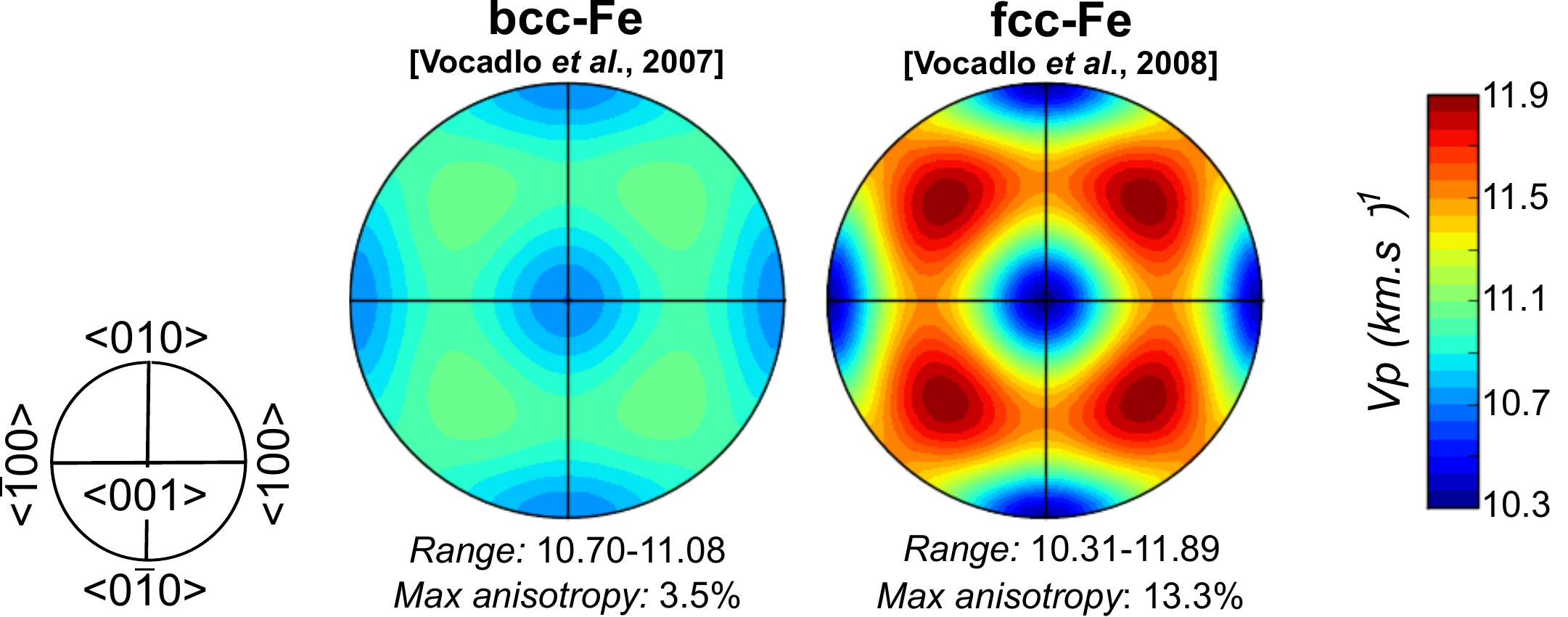}
\caption{P-wave velocity (in km/s) in bcc-Fe (left) and fcc-Fe (right) single crystal at inner core conditions ($\rho = 13155$~kg/m$^3$, $T=5500$~K) corresponding to single crystal elastic moduli (in GPa) of C$_{11} = 1505$, C$_{12} = 1160$ and C$_{44} = 256$ for bcc-Fe \citep{Vocadlo07} and C$_{11} = 1397$, C$_{12} = 1247$ and C$_{44} = 423$ for fcc-Fe \citep{Vocadlo08}. {Inlet on the right shows elastically equivalent directions in a cubic structure such as $\langle 100\rangle$, $\langle 010\rangle$, $\langle 001\rangle$, $\langle \overline{1}00\rangle$ and $\langle 0\overline{1}0\rangle$.}}
\label{figSingleCij}
\end{figure}

A each point of the grid, the local elastic tensor of the polycristal is then calculated by weighting the single crystal elastic moduli with the aggregate texture under the Hill approximation (Fig.~\ref{figVelocities}). Here, we use single-crystal elastic moduli from first-principles calculations (Fig.~\ref{figSingleCij}), with a 3.5\% and 13.3\% single crystal P-wave anisotropy for the bcc and fcc phase, respectively {\citep{Vocadlo07, Vocadlo08}}. In both cases, velocities are minimal and maximal along the $\langle 100 \rangle$ and $\langle 111 \rangle$ directions, respectively. Due to the cubic symmetry, velocity distributions in the single-crystal are complex and display 14 extremas, unlike hcp materials for which 3 extrema are found \citep{Lincot14} .

Velocity distributions, such as those presented in Fig.~\ref{figVelocities}, can be difficult to compare with actual observations of anisotropy. Hence, following the procedure of \citet{Lincot14}, we generate more than $100000$ synthetic seismic rays to probe the whole inner core and simulate seismic measurements. For each segment of the ray, we evaluate the slowness of the material by solving the Christoffel equation with the local elastic tensor deduced by linear interpolation of the elastic tensor of the aggregates located at the four nearest grid point of reference. For each ray, we estimate the normalized seismic travel times residual
\begin{equation}
\delta{t}/{t}  = \frac{s - s^{0}}{s^{0}}
\label{eqResidual}
\end{equation}
where $s$ is the simulated slowness of the seismic ray, and $s^0$ is the slowness of that same ray for an homogeneous and fully isotropic inner core.

\begin{figure*}
\centering
\caption{$P$-wave velocity (in km/s) in the present day inner core for geodynamical models \textit{Yos-Ran} (\textit{a,b}) and \textit{Yos-Strat} (\textit{c,d}) and either bcc-Fe  (\textit{a,c}) or fcc-Fe  (\textit{b,d}). Calculations are based on the textures of Fig.~\ref{figTextures} and the single crystal elastic moduli of Fig~\ref{figSingleCij}.}
\label{figVelocities}
\end{figure*}

\begin{figure}
\centering
\includegraphics[width=8cm]{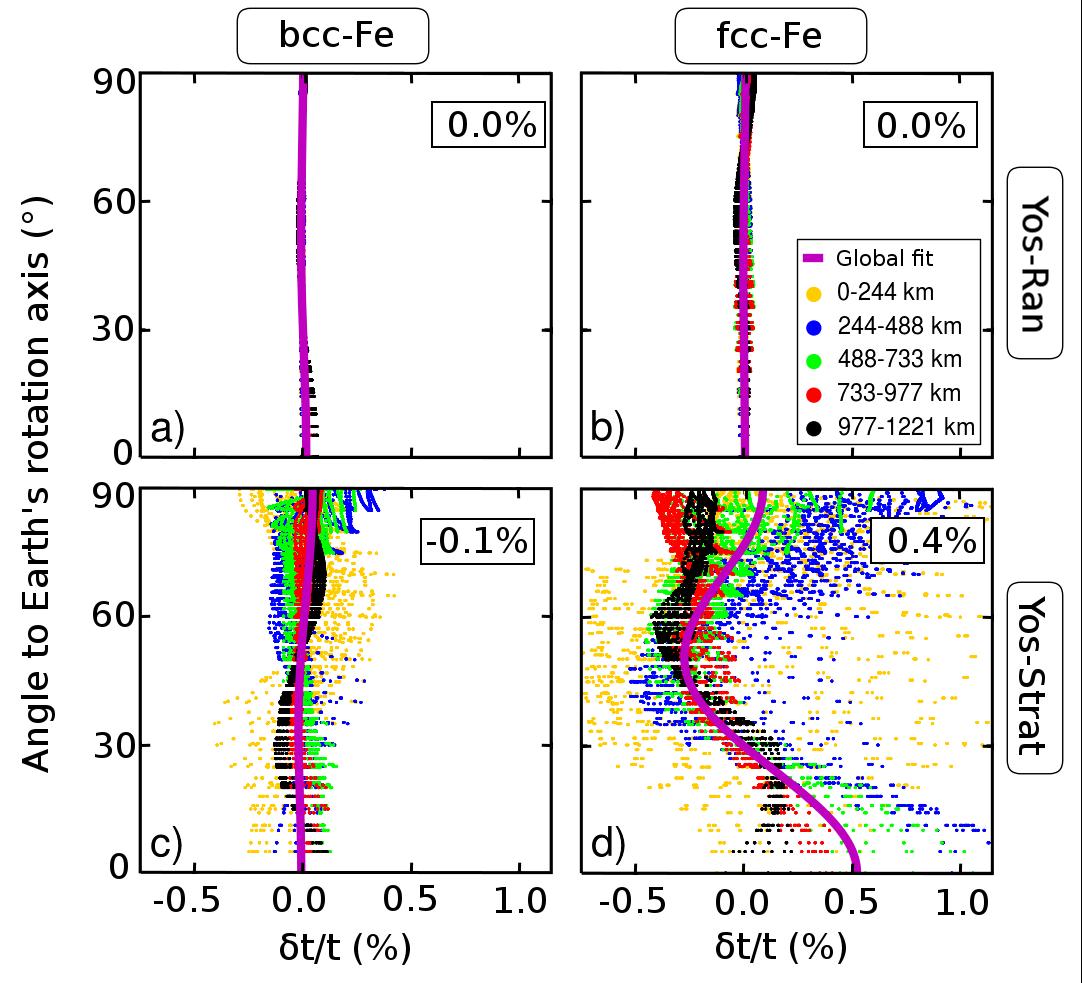}
\caption{Travel time residuals (dots) vs. angle between the ray and the Earth's rotation axis. Colors indicate the depth of turning point of the ray. Thick purple line is the fit of Eq.~\ref{eqIrving} to the data. Numbers in the insets indicate the global anisotropy for each model. Results plotted for geodynamical models \textit{Yos-Ran} (\textit{a,b}) and \textit{Yos-Strat} (\textit{a,b}) and either bcc-Fe  (\textit{a,c}) or fcc-Fe  (\textit{b,d}). Calculations based on the velocities of Fig.~4 and the single crystal elastic moduli of Fig~\ref{figSingleCij}.}
\label{figResiduals}
\end{figure}

Our model allows for a detailed analysis of anisotropy, including the depth and orientation
dependence of the travel time residuals (Fig.~\ref{figResiduals}). Those, however, can be difficult to compare with``global'' scale anisotropy reported in seismology publications. Assuming cylindrical symmetry of the inner core, \citet{Creager92} proposed to fit travel times residuals using
\begin{equation}
 \delta{t}/{t} = a + b\cos^{2}\zeta + c\cos^{4}\zeta,
 \label{eqIrving}
\end{equation}
where $a$, $b$ and $c$ are adjustable parameters and $\zeta$ is the angle between the ray and the Earth rotation axis. A measure of the global inner core anisotropy typically reported in the literature is the quantity $b + c$ \citep[e.g.][]{Irving11a}.

\section{Results for published elastic moduli}


Without stratification (\textit{Yos-Ran}, Figs.~\ref{figTextures}a,b), textures in the outer part of the inner core are nearly random while nearly axisymmetric textures develop in the central portion. This pattern is the result of the relatively slow deformation in the model of \citet{Yoshida96}, for which 100\% deformation is reached over the age of the inner core. Pole figures for bcc (Fig.~\ref{figTextures}a) and fcc (Fig.~\ref{figTextures}b) are inverted, with the 100 axes of the bcc phase preferentially aligned in the equatorial plane while, for fcc-Fe, they tend to lie in the North-South direction. This inversion of textures between bcc and fcc structures in well known in materials science and is due to an exchange of slip plane and direction between both structures.

With stratification (\textit{Yos-Strat}, Figs.~\ref{figTextures}c,d), strong textures are mostly concentrated at the top of the inner core, in a superficial layer where deformation is the largest \citep{Deguen09}.
In this superficial layer, the orientation of the texture varies strongly from mid latitude, where the horizontal stress is maximum (simple shear), to polar or equatorial regions for which deformation is similar to that of a corner flow. In opposition to model \textit{Yos-Ran}, most {of} the LPO developed in \textit{Yos-Strat} are not axisymmetric and heterogeneous,  both in direction and in intensity in the overall inner core.


Using the elastic moduli of Fig.~\ref{figSingleCij}, {anisotropy} in P-wave velocity is less then 0.1\% in model \textit{Yos-Ran}, whatever the location inside the inner core (Figs.~\ref{figVelocities}a,b). This also applies to the central portion of the inner core where the LPO is the strongest. This isotropy at the scale of the polycrystal is the result of the multiple symmetries for velocities in the single-crystal (Fig.~\ref{figSingleCij}) that average out at the scale of the aggregate. Note that this is an upper estimate that neglects the solid body rotation of the inner core that could occur during its formation.

The addition of stratified growth (model \textit{Yos-Strat}, Figs.~\ref{figVelocities}c,d) induces strong anisotropies in a limited region, at the base of the superficial layer at mid latitude. Anisotropy is more pronounced for the fcc phase (Fig.~\ref{figVelocities}d) reaching values of up to 10\% locally, close to that of the single crystal (13.3\%, Fig.~\ref{figSingleCij}b). Anisotropy for an inner core composed of bcc-Fe with the elastic moduli of Fig~\ref{figSingleCij} remains below 1.5\% (Fig.~\ref{figVelocities}c). In both cases, one should note that local anisotropy in model \textit{Yos-Strat} is complex, with numerous extrema that do not align with a North-South axisymmetry.


Fig.~\ref{figResiduals} presents the travel time residuals for seismic rays crossing the models of Fig.~\ref{figVelocities}. The \textit{Yos-Ran} model, without stratification, is isotropic, with a constant average ray velocity, whatever the direction of propagation or depth of turning point for the ray. Hence, the model of \citet{Yoshida96}, combined with a cubic phase of Fe and the elastic moduli of Fig~\ref{figSingleCij} is not consistent with observations of seismic anisotropy in the inner core.

With the addition of stratification, global anisotropy for bcc-Fe is low (0.1\%, Fig.~\ref{figResiduals}c) and opposite to that observed in seismic studies, i.e., North-South ray paths are slower than those in the equatorial plane. For fcc-Fe, a larger global anisotropy of 0.4\% (Fig.~\ref{figResiduals}d) is generated, as expected from Fig.~\ref{figVelocities}d. Nevertheless, note that the global anisotropy is a rather moderate in relation to that of the single crystal (Fig.~\ref{figSingleCij}) and the seismic reports of over 3\%  global anisotropy in the literature \citep{Irving11a}. The local maxima of $P$-wave anisotropy of over 10\% observed in Fig.~\ref{figVelocities}d are averaged out over the integration along the ray path.

Residuals in Fig.~\ref{figResiduals}d are minimal for rays with $\zeta = 60^\circ$, in opposition to actual seismic observations for which residual remain fairly constants for $\zeta > 45^\circ$ \citep{Irving11a}. The $C$-shape of the calculated residuals in model \textit{Yos-Strat} is mainly due to the contribution of rays with a turning point between 244 and 488~km below the ICB (blue dots),that is for rays that probe just below the first stratified layer with strong anisotropy in Figs.~\ref{figVelocities}c,d. This $C$-shape also implies that North-South and equatorial rays travel with a relatively fast speed. For deeper rays (red and black dots in Fig.~\ref{figResiduals}), the variations of the travel time anomalies are smaller with $\zeta$ (less than 0.5\%) but North-South rays travels faster than others. Also note that the fit of Eq.~\ref{eqIrving} is a poor representation of the travel time residuals and that the scattering of the data points around the fit is rather large (more than 1\% of $\delta{t}/{t}$, with a global anisotropy $b+c =0.4\%$).

In summary, we find that, within our synthetic grown inner core, no cubic phase of iron (bcc nor fcc) with the elastic moduli from first-principles calculations \citep{Vocadlo07,Vocadlo08} can generate a global anisotropy that compares with observations in seismology \citep[e.g.][]{Irving11a}.

\section{Generalisation to other elastic models}

Our conclusions from the previous section are strongly constrained by the choice of elastic moduli in Fig~\ref{figSingleCij}. This section intends to generalise our result to any cubic phase, whatever the choice of elastic moduli. To that end,
we introduce a Monte-Carlo approach to determine the effect of single-crystal elasticity on the global anisotropy at inner core scale.

We select 3000 random sets of elastic moduli constrained by conditions for mechanical stability \citep{wallace1972}
\begin{equation}
C_{11} - C_{12} >0,  C_{11} + 2 C_{12} >0, C_{44}>0.
\end{equation}
and an average bulk and shear moduli in the Hill average that match those of PREM \citep[K = 1384~GPa, G = 166~GPa][]{PREM} with 15\%. These 3000 sets of  elastic moduli   sample all possibilities for single-crystal elasticity for cubic Fe at inner core conditions.

Single-crystal anisotropy can be quantified using the dimensionless parameter $K$, as defined in \citet{Karki97}:
\begin{equation}
K = \frac{2C_{44} + C_{12}}{C_{11}} -1
\label{eq:karki}
\end{equation}
For an isotropic crystal, $K = 0$. $K$ becomes positive (negative) when P-wave velocities are minimal (maximal) along $\langle 100\rangle$. With the elastic moduli of Fig.~\ref{figSingleCij}, $K$ is positive.

\begin{figure}
\centering
\includegraphics[width=8cm]{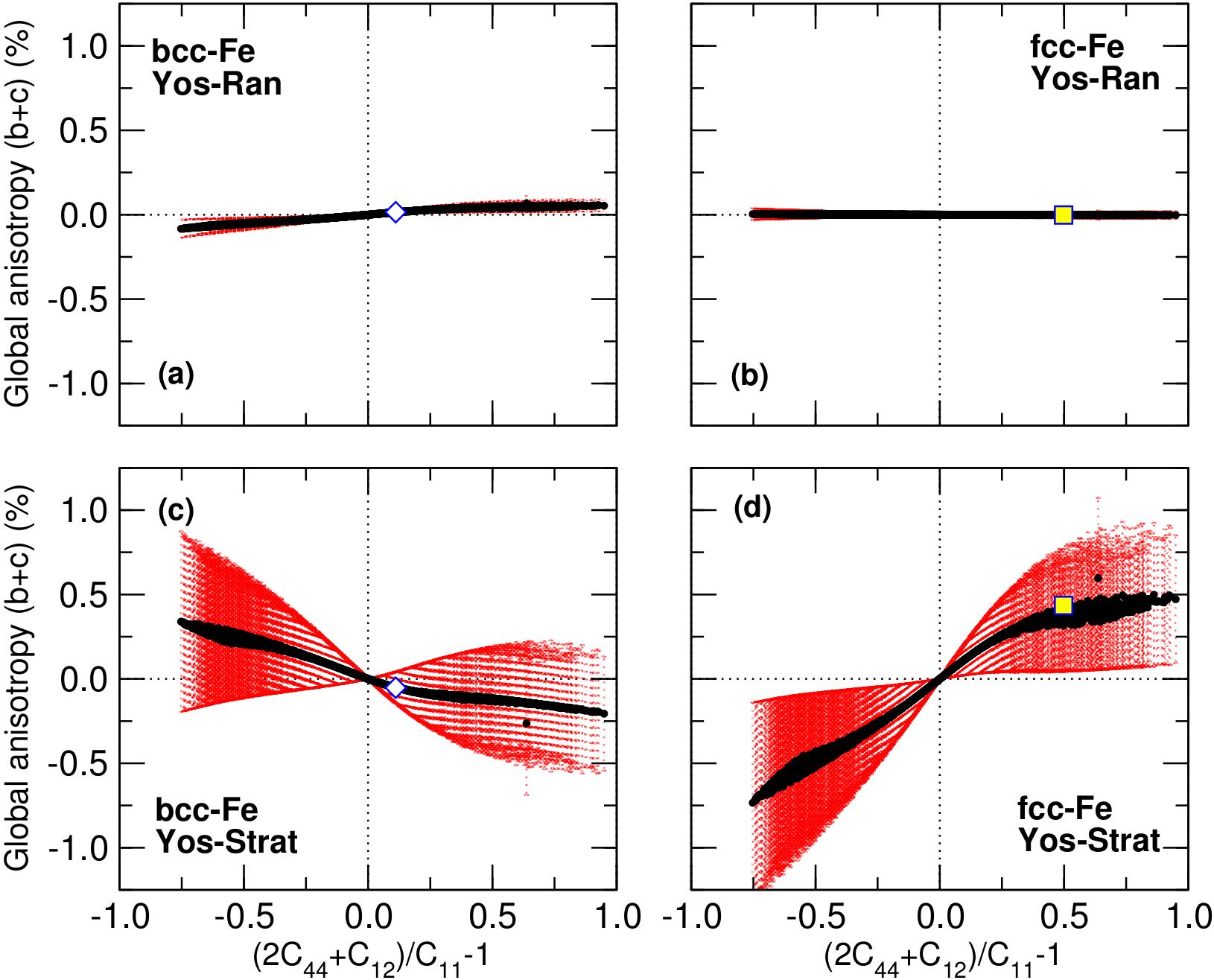}
\caption{Global inner core anisotropy vs. single crystal anisotropy for  geodynamical models \textit{Yos-Ran} (\textit{a,b}) and \textit{Yos-Strat} (\textit{c,d}) and either bcc-Fe  (\textit{a,c}) or fcc-Fe  (\textit{b,d}). For each figure, calculations were performed on 3000 random sets of elastic moduli covering all possible values for the inner core (solid black dots, see text). The error range (red area) is an estimate of the spread of residuals for individual rays. White-filled diamond and yellow-filled squares are the results of calculations based on the elastic moduli of Fig~\ref{figSingleCij}.  }
\label{figMonteCarlo}
\end{figure}

For each of the 3000 sets of elastic moduli, we repeat the procedure above to characterize the global inner core anisotropy for each of the 4 models, i.e. geodynamical models \textit{Yos-Ran} and \textit{Yos-Strat}, with an inner core consisting of either a bcc- or an fcc-structured material. Global anisotropy $(b+c)$ plotted as a function of the dimensionless parameter $K$ follows a linear trend (Fig.~\ref{figMonteCarlo}): global anisotropy in the inner core increases with single crystal anisotropy.

Without stratification, no cubic material can produce sufficient global anisotropy to match seismic observations. This result is general and does not depend on the crystal structure, nor the set of elastic moduli. Stratification enhances global inner core anisotropy (model \textit{Yos-Strat}, Fig.~\ref{figMonteCarlo}c,d) but global anisotropy does not exceed ~0.5\%, well below seismic observations. {Anisotropies} are larger for the case of fcc structure.

Seismic observations of inner core anisotropy are heterogeneous, depth, and geographically dependent \citep{Irving11a}. Hence, a global fit based on Eq.~\ref{eqIrving} on 100000 random rays may not be relevant for comparing with seismic observations. Hence, for each of the 3000 random sets of elastic moduli, Fig.~\ref{figMonteCarlo} also presents the standard deviation of all individual residuals. Using \textit{Yos-Ran} model, none of the residuals will ever exceed 0.2\% whatever the elastic moduli, for all ray paths. For the \textit{Yos-Strat} model, standard deviations of all residuals can reach values of up to 1\% for extremely anisotropic single crystal elastic moduli. Hence, seismic observations of above 3\% anisotropy are incompatible with our model, whatever the ray path, whatever the elastic moduli, and whatever the choice of crystal structure, bcc or fcc. If the inner core is cubic Fe, its seismic anisotropy cannot be due to plastic deformation alone.

\section{Conclusions}

We used the framework of \cite{Lincot14} to study whether inner core anisotropy arises from plastic deformation of a cubic phase. Our analysis relies on a geodynamical model derived from that of \citet{Yoshida96}, Fe alloy either in the bcc or fcc structure deforming along classical slip systems, and a Monte-Carlo approach for probing all possible sets of elastic moduli.

Within our assumptions, no cubic phase of Fe can produce a global inner core anisotropy larger than 0.5\%, far below seismic observations. This observation results from the multiple symmetries over multiple scales involved: symmetries in single-crystal P-wave velocities and multiple equivalent slip systems combined with large inner core portions probed with each seismic ray.

Hence, the observed inner core anisotropy does not arise from deformation-induced plastic deformation of a cubic phase of an Fe alloy. If the inner core material is indeed cubic, other mechanisms for generating anisotropy are required and preserved, such as solidification texturing, grain growths or melt inclusions.  In any case, anisotropic structures in the inner core should be coherent over large scales for a cubic material to be compatible with observations of seismic anisotropy.



\begin{acknowledgments}
The authors wish to thank M. Bergman and an anonymous reviewer for constructive comments and the Geodynamo group at ISTerre for useful discussions.
This work has been financed by the program PNP of CNRS/INSU and labex OSUG@2020. SM received support from the Institut Universitaire de France.
Calculations were made at Centre de Calcul Commun of the OSUG.
\end{acknowledgments}

%


\begin{thebibliography}{27}
\providecommand{\natexlab}[1]{#1}
\expandafter\ifx\csname urlstyle\endcsname\relax
  \providecommand{\doi}[1]{doi:\discretionary{}{}{}#1}\else
  \providecommand{\doi}{doi:\discretionary{}{}{}\begingroup
  \urlstyle{rm}\Url}\fi

\bibitem[{\textit{Belonoshko et~al.}(2008)\textit{Belonoshko, Skorodumova,
  Rosengren, and Johansson}}]{belonoshko2008science}
Belonoshko, A.~B., N.~V. Skorodumova, A.~Rosengren, and B.~Johansson, Elastic
  anisotropy of {E}arth's inner core, \textit{Science}, \textit{319}, 797--800,
  \doi{10.1126/science.1141374}, 2008.

\bibitem[{\textit{Bergman}(1997)}]{Bergman97}
Bergman, M.~I., Measurements of elastic anisotropy due to solidification
  texturing and the implications for the {E}arth's inner core, \textit{Nature},
  \textit{389}, 60--63, \doi{10.1038/37962}, 1997.

\bibitem[{\textit{Bouchet et~al.}(2013)\textit{Bouchet, Mazevet, Morard, Guyot,
  and Musella}}]{bouchet2013prb}
Bouchet, J., S.~Mazevet, G.~Morard, F.~Guyot, and R.~Musella, Ab initio
  equation of state of iron up to 1500~{GPa}, \textit{Phys. Rev. B},
  \textit{87}, {094102}, \doi{10.1103/PhysRevB.87.094102}, 2013.

\bibitem[{\textit{{Creager}}(1992)}]{Creager92}
{Creager}, K.~C., {Anisotropy of the inner core from differential travel times
  of the phases PKP and PKIKP}, \textit{Nature}, \textit{356}, 309--314,
  \doi{10.1038/356309a0}, 1992.

\bibitem[{\textit{Deguen}(2012)}]{deguen2012epsl}
Deguen, R., Structure and dynamics of {E}arth's inner core, \textit{Earth Planet.
  Sci. Lett.}, \textit{333-334}, 211--225, \doi{10.1016/j.epsl.2012.04.038},
  2012.

\bibitem[{\textit{{Deguen} and {Cardin}}(2009)}]{Deguen09}
{Deguen}, R., and P.~{Cardin}, Tectonic history of the {E}arth's inner core
  preserved in its seismic structure, \textit{Nature. Geosci.}, \textit{2},
  419--422, \doi{10.1038/ngeo522}, 2009.

\bibitem[{\textit{Deguen et~al.}(2011)\textit{Deguen, Cardin, Merkel, and
  A.}}]{Deguen11a}
Deguen, R., P.~Cardin, S.~Merkel, and L.~R. A., Texturing in {E}arth's inner
  core due to preferential growth in its equatorial belt, \textit{Phys. Earth
  Planet. Inter.}, \textit{188}, 173--184, \doi{10.1016/j.pepi.2011.08.008},
  2011.

\bibitem[{\textit{Deuss}(2014)}]{deuss2014areps}
Deuss, A., Heterogeneity and anisotropy of {E}arth's inner core, \textit{Ann.
  Rev. Earth Planet. Sc.}, \textit{42}, 103--126,
  \doi{10.1146/annurev-earth-060313-054658}, 2014.

\bibitem[{\textit{Dziewo\'{n}ski and Anderson}(1981)}]{PREM}
Dziewo\'{n}ski, A.~M., and D.~L. Anderson, Preliminary reference {E}arth model,
  \textit{Phys. Earth Planet. Inter.}, \textit{25}, 297--356,
  \doi{10.1016/0031-9201(81)90046-7}, 1981.

\bibitem[{\textit{Irving and Deuss}(2011{\natexlab{a}})}]{Irving11a}
Irving, J. C.~E., and A.~Deuss, Hemispherical structure in inner core velocity
  anisotropy, \textit{J. Geophys. Res.}, \textit{116}, {B04307},
  \doi{10.1029/2010JB007942}, 2011{\natexlab{a}}.

\bibitem[{\textit{Irving and Deuss}(2011{\natexlab{b}})}]{Irving11b}
Irving, J. C.~E., and A.~Deuss, Stratified anisotropic structure at the top of
  Earth's inner core: A normal mode study, \textit{Phys. Earth Planet. Inter.},
  \textit{186}, \doi{10.1016/j.pepi.2011.03.003}, 2011{\natexlab{b}}.

\bibitem[{\textit{Karki et~al.}(1997)\textit{Karki, Stixrude, Clark, Warren,
  Ackland, and Crain}}]{Karki97}
Karki, B., L.~Stixrude, S.~Clark, M.~Warren, G.~Ackland, and J.~Crain,
  Structure and elasticity of {MgO} at high pressure, \textit{Am. Mineral.},
  \textit{82}, 51--60, 1997.

\bibitem[{\textit{Kocks et~al.}(1998)\textit{Kocks, Tom\'e, and
  Wenk}}]{kocks1998}
Kocks, U.~F., C.~N. Tom\'e, and H.~R. Wenk, \textit{Texture and Anisotropy:
  Preferred Orientations and their Effects on Material Properties}, Cambridge
  Univ. Press, Cambridge, 1998.

\bibitem[{\textit{Lebensohn and Tom\'e}(1993)}]{Lebensohn93}
Lebensohn, R.~A., and C.~N. Tom\'e, Self-consistent anisotropic approach for
  the simulation of plastic deformation and texture development of
  polycrystals: application to zirconium alloys., \textit{Acta Metall. Mater.},
  \textit{41}, 2611--2624, 1993.

\bibitem[{\textit{Lincot et~al.}(2014)\textit{Lincot, Deguen, Merkel, and
  Cardin}}]{Lincot14}
Lincot, A., R.~Deguen, S.~Merkel, and P.~Cardin, Seismic response and
  anisotropy of a model hcp iron inner core, \textit{C. R. Geosci.},
  \textit{346}, 148--157, \doi{10.1016/j.crte.2014.04.001}, 2014.

\bibitem[{\textit{Lythgoe et~al.}(2014)\textit{Lythgoe, Deuss, Rudge, and
  Neufeld}}]{Lythgoe14}
Lythgoe, K.~H., A.~Deuss, J.~F. Rudge, and J.~A. Neufeld, Earth's inner core:
  Innermost inner core or hemispherical variations?, \textit{Earth Planet. Sci.
  Lett.}, \textit{385}, 181--189, \doi{{10.1016/j.epsl.2013.10.049}}, 2014.

\bibitem[{\textit{Mikhaylushkin et~al.}(2007)\textit{Mikhaylushkin, Simak,
  Dubrovinsky, Dubrovinskaia, Johansson, and Abrikosov}}]{mikhaylushkin2007prl}
Mikhaylushkin, A.~S., S.~I. Simak, L.~Dubrovinsky, N.~Dubrovinskaia,
  B.~Johansson, and I.~A. Abrikosov, Pure iron compressed and heated to extreme
  conditions, \textit{Phys. Rev. Lett.}, \textit{99}, {165505},
  \doi{10.1103/PhysRevLett.99.165505}, 2007.

\bibitem[{\textit{Modak et~al.}(2007)\textit{Modak, Verma, Rao, Godwal,
  Stixrude, and Jeanloz}}]{modak2007jpcm}
Modak, P., A.~K. Verma, R.~S. Rao, B.~K. Godwal, L.~Stixrude, and R.~Jeanloz,
  Stability of the hcp phase and temperature variation of the axial ratio of
  iron near {E}arth-core conditions, \textit{J. Phys.: Condens. Matter},
  \textit{19}, {016208}, \doi{10.1088/0953-8984/19/1/016208}, 2007.

\bibitem[{\textit{Morard et~al.}(2014)\textit{Morard, Andrault, Antonangeli,
  and Bouchet}}]{morard2014crg}
Morard, G., D.~Andrault, D.~Antonangeli, and J.~Bouchet, Properties of iron
  alloys under the {E}arth's core conditions, \textit{C R Geosci},
  \textit{346}, 130--139, \doi{10.1016/j.crte.2014.04.007}, 2014.

\bibitem[{\textit{{Singh} et~al.}(2000)\textit{{Singh}, {Taylor}, and
  {Montagner}}}]{Singh00}
{Singh}, S.~C., M.~A.~J. {Taylor}, and J.~P. {Montagner}, On the presence of
  liquid in {E}arth's inner core, \textit{Science}, \textit{287}, 2471--2474,
  \doi{10.1126/science.287.5462.2471}, 2000.

\bibitem[{\textit{{Souriau}}(2003)}]{Souriau03}
{Souriau}, A., {The seismological picture of the inner core: structure and
  rotation}, \textit{C. R. Geosci.}, \textit{335}, 51--63,
  \doi{10.1016/S1631-0713(03)00010-5}, 2003.

\bibitem[{\textit{Tanaka}(2012)}]{Tanaka12}
Tanaka, S., Depth extent of hemispherical inner core from {PKP(DF)} and
  {PKP(Cdiff)} for equatorial paths, \textit{Phys. Earth Planet. Inter.},
  \textit{210}, 50--62, \doi{10.1016/j.pepi.2012.08.001}, 2012.

\bibitem[{\textit{Tateno et~al.}(2010)\textit{Tateno, Hirose, Ohishi, and
  Tatsumi}}]{Tateno10}
Tateno, S., K.~Hirose, Y.~Ohishi, and Y.~Tatsumi, The structure of iron in
  {E}arth's inner core, \textit{Science}, \textit{330}, 359--361,
  \doi{10.1126/science.1194662}, 2010.

\bibitem[{\textit{Vo\v{c}adlo}(2007)}]{Vocadlo07}
Vo\v{c}adlo, L., Ab initio calculations of the elasticity of iron and iron
  alloys at inner core conditions: Evidence for a partially molten inner core?,
  \textit{Earth Planet. Sci. Lett.}, \textit{254}, 227--232,
  \doi{10.1016/j.epsl.2006.09.046}, 2007.

\bibitem[{\textit{Vo\v{c}adlo et~al.}(2008)\textit{Vo\v{c}adlo, Wood, Alf\`e,
  and Price}}]{Vocadlo08}
Vo\v{c}adlo, L., I.~G. Wood, D.~Alf\`e, and G.~D. Price, Ab initio calculations
  on the free energy and high {P-T} elasticity of face-centred-cubic iron,
  \textit{Earth Planet. Sci. Lett.}, \textit{268}, 444--449,
  \doi{10.1016/j.epsl.2008.01.043}, 2008.

\bibitem[{\textit{Wallace}(1972)}]{wallace1972}
Wallace, D.~C., \textit{Thermodynamics of Crystals}, Wiley, New York, 1972.

\bibitem[{\textit{{Yoshida} et~al.}(1996)\textit{{Yoshida}, {Sumita}, and
  {Kumazawa}}}]{Yoshida96}
{Yoshida}, S., I.~{Sumita}, and M.~{Kumazawa}, {Growth model of the inner core
  coupled with the outer core dynamics and the resulting elastic anisotropy},
  \textit{J. Geophys. Res.}, \textit{101}, 28085--28104,
  \doi{10.1029/96JB02700}, 1996.

\end{thebibliography}

\end{article}
\end{document}